\documentclass[10pt,prd,aps,twocolumn,showpacs,showkeys,superscriptaddress,groupedaddress,floatfix]{revtex4-1}
\usepackage{amsmath}
\usepackage{amsfonts}
\usepackage{amssymb}
\usepackage{graphicx} 
\usepackage{mathrsfs}
\usepackage{url}
\usepackage[english]{babel}

\def\ee{\mbox{$\left(e,e^{\prime}\right)$\ }}
\def\eep{\mbox{$\left(e,e^{\prime}p\right)$\ }}

\begin{document}
\title{Relativistic descriptions of final-state interactions
in charged-current neutrino-nucleus scattering at ArgoNeuT
kinematics}

\author{Andrea Meucci}
\author{Carlotta Giusti}
\author{Matteo Vorabbi}
\affiliation{Dipartimento di Fisica, 
Universit\`{a} degli Studi di Pavia and \\
INFN,
Sezione di Pavia, via A. Bassi 6, I-27100 Pavia, Italy}
\begin{abstract}
The analysis of the recent charged-current neutrino-nucleus
scattering cross sections measured by the ArgoNeuT Collaboration requires relativistic 
theoretical descriptions also accounting for the role of final-state 
interactions. In this work, we evaluate differential neutrino-nucleus
cross sections with  
the relativistic Green's function model, where final-state interactions are described in 
the inclusive scattering consistently with the exclusive scattering using a complex 
optical potential.
The sensitivity to the parameterization adopted for the 
phenomenological optical potential is discussed. The predictions of the 
relativistic Green's function model are compared with the results of
different descriptions of final-state interactions.
\end{abstract}
\date{\today}
\maketitle 


\section{Introduction}
\label{intro}

In the past decade many neutrino oscillation results have been presented by 
different collaborations \cite{superkam,superkam11,icarus13,sno,minos12,t2k11,minibPhysRevLett13,
minibPhysRevD12,PhysRevD.86.052009,PhysRevLett.108.131801,PhysRevLett.108.171803,
An:2012bu,PhysRevLett.108.191802,PhysRevD.74.072003,PhysRevD.64.112007},
and a phenomenological extension of the Standard Model has been proposed that involves three 
neutrino mass states, over which the three
flavors of neutrinos are distributed. Despite its successful
predictions, this can be considered as an extension of the Standard Model
that does not address fundamental questions, e.g., small masses and large mixing angles
compared to quark sector, and has raised a large debate over other possible unexpected properties of 
neutrinos that could lead to a more complete understanding of neutrino 
physics\ \cite{Conrad:2012qt,Petcov:2013poa,PhysRevD.85.083522,PhysRevD.83.073006,
1475-7516-2011-09-034,PhysRevD.70.073004}.
To gain a deeper understanding of neutrino phenomenology
the reduction of uncertainties in baseline neutrino oscillation experiments is mandatory.

Because of the interest in oscillation measurements, 
in recent years various neutrino-nucleus differential cross sections  
have been presented \cite{miniboone,miniboonenc,miniboone-ant,Nakajima:2010fp,argoneut} 
and 
are planned in the near future~\cite{minibooneweb,minerva,t2k}. 
Differential cross sections 
are important to obtain a complete kinematical determination of neutrino-nucleus scattering 
and a clear understanding of neutrino-nucleus reactions is crucial for the analysis 
of experimental measurements. 
  
The ArgoNeuT Collaboration has recently reported~\cite{argoneut} a 
measurement of the muon neutrino charged-current (CC) flux-averaged 
differential cross section on $^{40}$Ar in an energy range up to 50 GeV. 
A liquid Argon detector is very interesting because it
has excellent potentialities to make precise measurements of a very large class of 
neutrino interactions from the MeV energy scale to multi-GeV 
events \cite{Karagiorgi:2013cwa}. 
The ArgoNeuT measurement has proven the validity of this experimental technique 
and, hoperfully, new data will be available in the future. For instance, a calculation of 
neutrino capture cross sections for solar neutrinos  that could be addressed by this 
new generation of detectors is presented in \cite{PhysRevC.87.014607}.

The energy region 
considered in the ArgoNeuT experiment, with average neutrino energy of   
$\langle E_{\nu}\rangle = 4.3 $ GeV, requires the use of a relativistic model, where 
not only relativistic kinematics 
is considered, but also nuclear dynamics and current operators 
are described within a relativistic framework. 

The first measurement of the charged-current quasielastic (CCQE)
flux-averaged double-differential muon neutrino cross
section on $^{12}$C by the MiniBooNE 
Collaboration  \cite{miniboone} has raised  extensive
discussions. In particular, the experimental cross section is usually 
 underestimated by the
relativistic Fermi gas  model and by other more sophisticated models 
based on the impulse approximation 
\cite{Benhar:2010nx,Benhar:2011wy,Butkevich:2010cr,Butkevich:2011fu,jusz10}, unless 
the nucleon axial mass  
$M_A$ is significantly enlarged with respect to the world average value of 1.03 GeV/$c^2$. 
It is reasonable to assume that the larger axial mass obtained from the 
MiniBooNE data on $^{12}$C can  be interpreted as an effective way to include medium effects 
that are not taken into account by the models; this is another indication 
that a precise knowledge of lepton-nucleus cross 
sections, where uncertainties on nuclear effects are reduced as much as 
possible, is necessary.
Moreover, 
any model aimed to describe neutrino-nucles scattering should first be 
tested against electron scattering data in the same kinematic region.

At intermediate energy, quasielastic (QE) electron scattering 
calculations~\cite{Boffi:1993gs,book}, which were able to
successfully describe a wide number of experimental data, can provide a 
useful tool to study neutrino-induced processes. 
There are, however, indications that the reaction can have significant contributions 
from effects beyond  the impulse approximation (IA) in some energy regions 
where the neutrino flux has still significant  strength.  
For instance, in the models of \cite{Martini:2009uj,Martini:2010ex,
Martini:2011wp,Nieves:2011pp,Nieves:2011yp,Nieves201390}
the contribution of multinucleon excitations to CCQE scattering has been 
found sizable and able to bring the theory in agreement with the experimental
MiniBooNE cross sections without increasing the value of $M_A$. 
The role of processes involving two-body currents compared to the IA models
has been discussed in \cite{Benhar:2011wy,Amaro:2010sd,Amaro:2011qb,AmaroAntSusa,
bodek11,Golan:2013jtj}. 
A careful evaluation of all nuclear effects 
and of the relevance of multinucleon emission and of some non-nucleonic 
contributions~\cite{PhysRevC.79.034601,Leitner:2010kp,
PhysRevC.83.054616,FernandezMartinez2011477} would be, without a doubt, useful for a deeper
understanding of the reaction dynamics.  

The relevance of final state interactions (FSI) has been clearly stated for the exclusive 
$\eep$ reaction, where the use of a
complex optical potential (OP) in the distorted-wave impulse approximation (DWIA) 
is required~\cite{Boffi:1993gs,book,Udias:1993xy,Meucci:2001qc,
Meucci:2001ja,Meucci:2001ty,Radici:2003zz,Giusti:2011it}. The imaginary part 
of the OP produces an absorption that reduces
the cross section and accounts for the loss of part of the incident flux 
in the elastically scattered beam to 
the inelastic channels which are open.
In the inclusive scattering only the emitted lepton is detected,
the final nuclear state is not determined and all
elastic and inelastic channels contribute. Thus, a different treatment of FSI is
required, where all final-state channels are retained
and the total flux, although redistributed among all possible
channels, is conserved.

Different approaches have been used to describe FSI in relativistic calculations for 
the inclusive QE electron- and neutrino-nucleus 
scattering~\cite{Maieron:2003df,Caballero:2006wi,Caballero:2009sn,Meucci:2003cv,
Meucci:2003uy,Meucci:2004ip,Meucci:2006cx,Meucci:2006ir,Meucci:2008zz,Giusti:2009sy,
Meucci:2009nm,Meucci:2011pi,refId0}. 
In the relativistic plane-wave impulse approximation (RPWIA), FSI 
are simply neglected. In other approaches FSI are included in DWIA 
calculations where the 
final nucleon state is evaluated with real potentials, either retaining 
only the real part of the relativistic energy-dependent complex optical potential 
(rROP), or using the same relativistic mean field potential considered in 
describing the initial nucleon state (RMF). 
Although conserving the flux, the rROP is unsatisfactory from a theoretical 
point of view.
On the contrary,  the RMF, where the same strong energy-independent real 
potential is used for both bound and scattering states, fulfills the 
dispersion relations~\cite{hori} and also the continuity equation. 

In a different description of FSI relativistic Green's function (RGF) 
techniques~\cite{Capuzzi:1991qd,Meucci:2003cv,Meucci:2003uy,Capuzzi:2004au,Meucci:2005pk,
Meucci:2009nm,Meucci:2011pi,Giusti:cortona11,esotici2}
are used. 
In the RGF model, under suitable approximations, which are basically related 
to the IA, the components of the nuclear response are written 
in terms of the single particle optical model Green's function;
its spectral representation, that is
based on a biorthogonal expansion in terms of a non-Hermitian
optical potential $\cal H$ and of its Hermitian conjugate $\cal H^{\dagger}$, can be exploited 
to avoid the explicit calculation of the single particle Green's function and obtain the 
components of the hadron tensor \cite{Meucci:2003uy,Meucci:2003cv}. 
Calculations require matrix elements 
of the same type as the DWIA ones
of the exclusive \eep process in 
\cite{Meucci:2001qc}, but involve 
eigenfunctions of both $\cal H$ and $\cal H^{\dagger}$, where the imaginary 
part has an opposite sign and gives in one case a loss and in the 
other case a gain of strength. The RGF formalism allows us to reconstruct the 
flux lost into nonelastic channels in the case of the inclusive response 
starting from the complex optical potential which describes elastic 
nucleon-nucleus scattering data and to include contributions which are not 
included in the RMF and in other models based on the IA. Moreover, with the use
of the same complex optical potential, it provides a consistent treatment of 
FSI in the exclusive and in the inclusive scattering. In addition, 
because of the analyticity properties of the optical potential, 
it fulfills the Coulomb sum rule~\cite{hori,Capuzzi:1991qd,Meucci:2003uy}.

These different descriptions of FSI have been compared in~\cite{Meucci:2009nm} 
for the inclusive QE electron scattering, in \cite{Meucci:2011pi} for 
the CCQE neutrino scattering, and in \cite{Meucci:2011vd,Meucci:ant,Meucci:2011nc} 
with the CCQE and NCE MiniBooNE data. Both
RMF and RGF are able to describe the shape of the CCQE experimental 
data, only the RGF gives cross sections of the same magnitude as 
the experimental ones without the need to 
increase the value of $M_A$~\cite{Meucci:2011vd,Meucci:ant}. 
Similar results are obtained in \cite{Meucci:2011nc}, where the RGF 
results and their interpretation in comparison with the NCE data from MiniBooNE 
are discussed. 

In this paper the results of different relativistic 
descriptions of FSI for CC $\nu$-nucleus scattering are presented and 
discussed 
for the differential cross section averaged over the  $\nu_{\mu}$ ArgoNeuT 
flux. 
We are aware of the interpretative questions that may be connected to
the use of models
developed for the QE regime in a kinematic situation, with  
the $\nu_{\mu}$ ArgoNeuT flux up to $50$ GeV, 
where other processes beyond the IA, which are not included in the models
considered here, can give significant contributions. Nevertheless
we believe that our calculations can give useful information about the
role of nuclear effects in the analysis of $\nu_{\mu}-^{40}$Ar scattering and
about the 
uncertainties which are related to their evaluation.

\section{Results and discussion}
\label{results} 

\begin{figure}
\includegraphics[scale=.37]{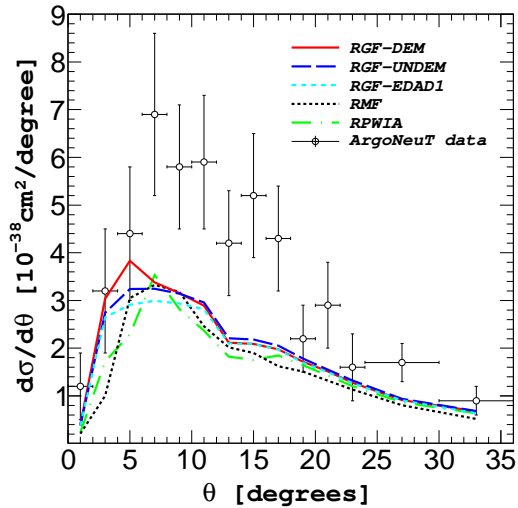} 
\vskip -0.2cm
\caption{(Color online) Flux-averaged differential 
cross section  $d\sigma/ d\theta$  for the reaction $^{40}$Ar$\left( \nu_{\mu}, \mu^- \right)$ 
 as a function of the muon scattering angle $\theta$.
 The data are
from ArgoNeuT \cite{argoneut}.
\label{csteta} }
\end{figure}
\begin{figure}
\includegraphics[scale=.37]{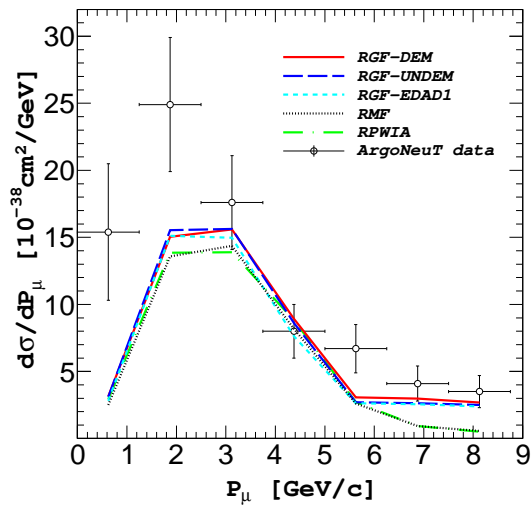} \vskip -0.2cm
\caption{(Color online) Flux-averaged differential 
cross section  $d\sigma/ dP_{\mu}$  for the reaction $^{40}$Ar$\left( \nu_{\mu}, \mu^- \right)$ 
 as a function of the muon momentum $P_{\mu}$.
 The data are
from ArgoNeuT \cite{argoneut}.
\label{cspmu} }
\end{figure}
\begin{figure}
\includegraphics[scale=.37]{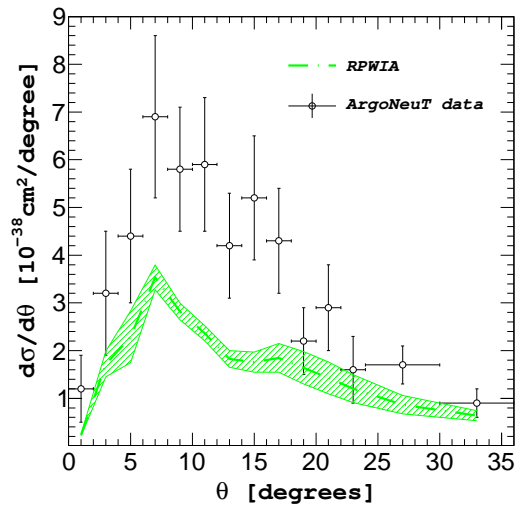} \\
\includegraphics[scale=.37]{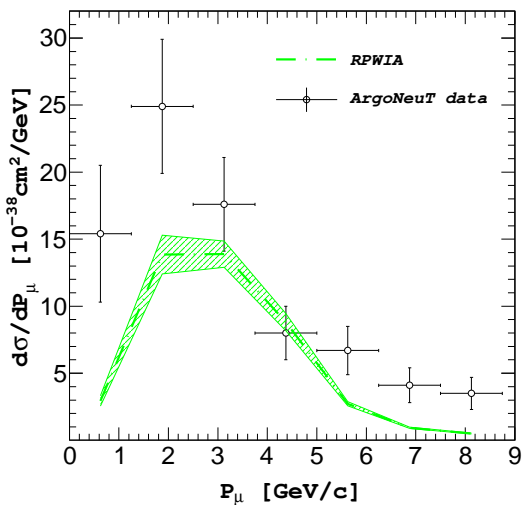}
\vskip -0.2cm
\caption{(Color online) Estimated uncertainties for the flux-averaged differential 
cross sections  $d\sigma/ d\theta$ and $d\sigma/ dP_{\mu}$ on  $^{40}$Ar 
calculated in RPWIA.
 The data are
from ArgoNeuT \cite{argoneut}.
\label{csrpwia-area} }
\end{figure}
\begin{figure}
\includegraphics[scale=.37]{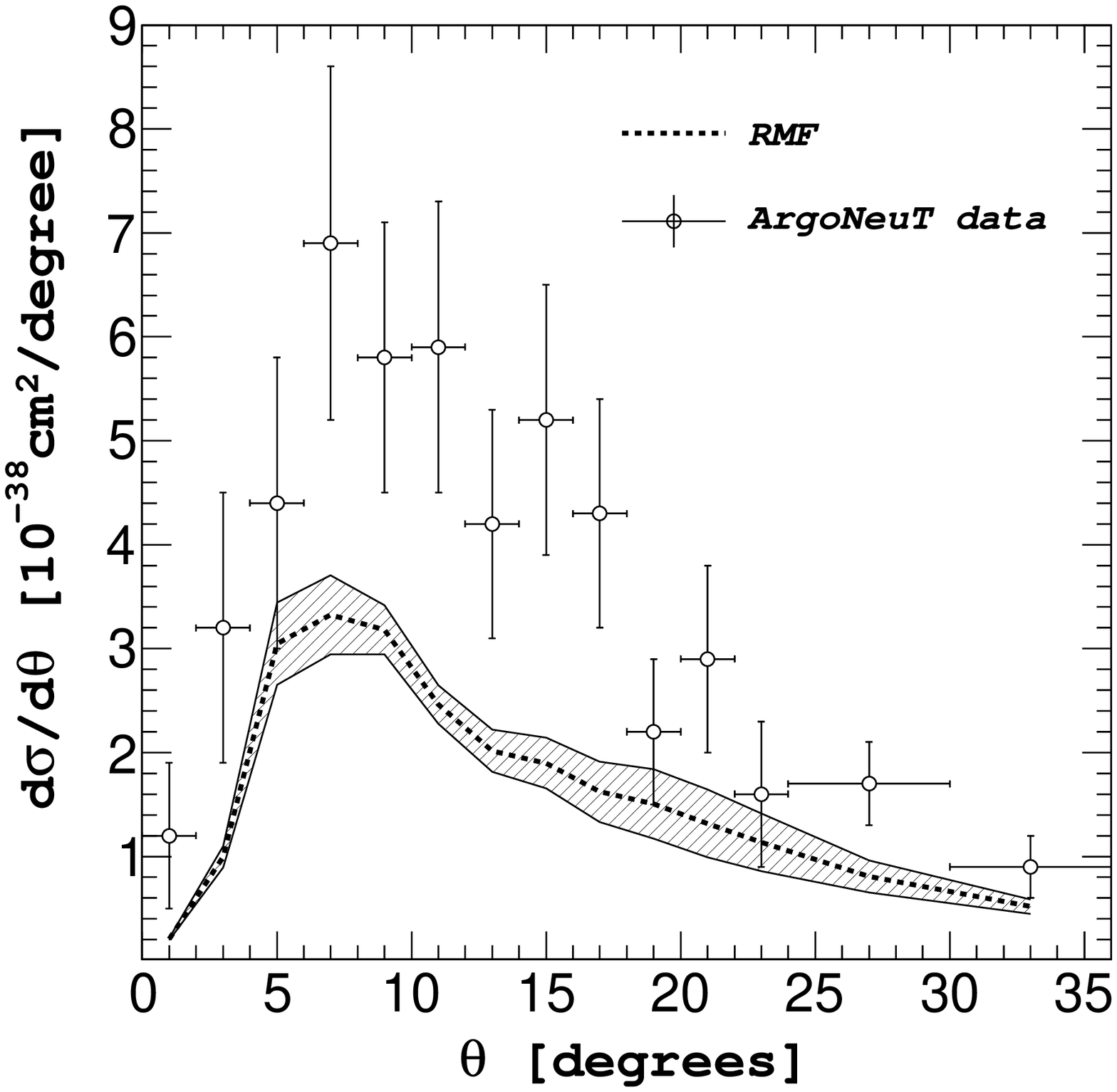}  \\
\includegraphics[scale=.37]{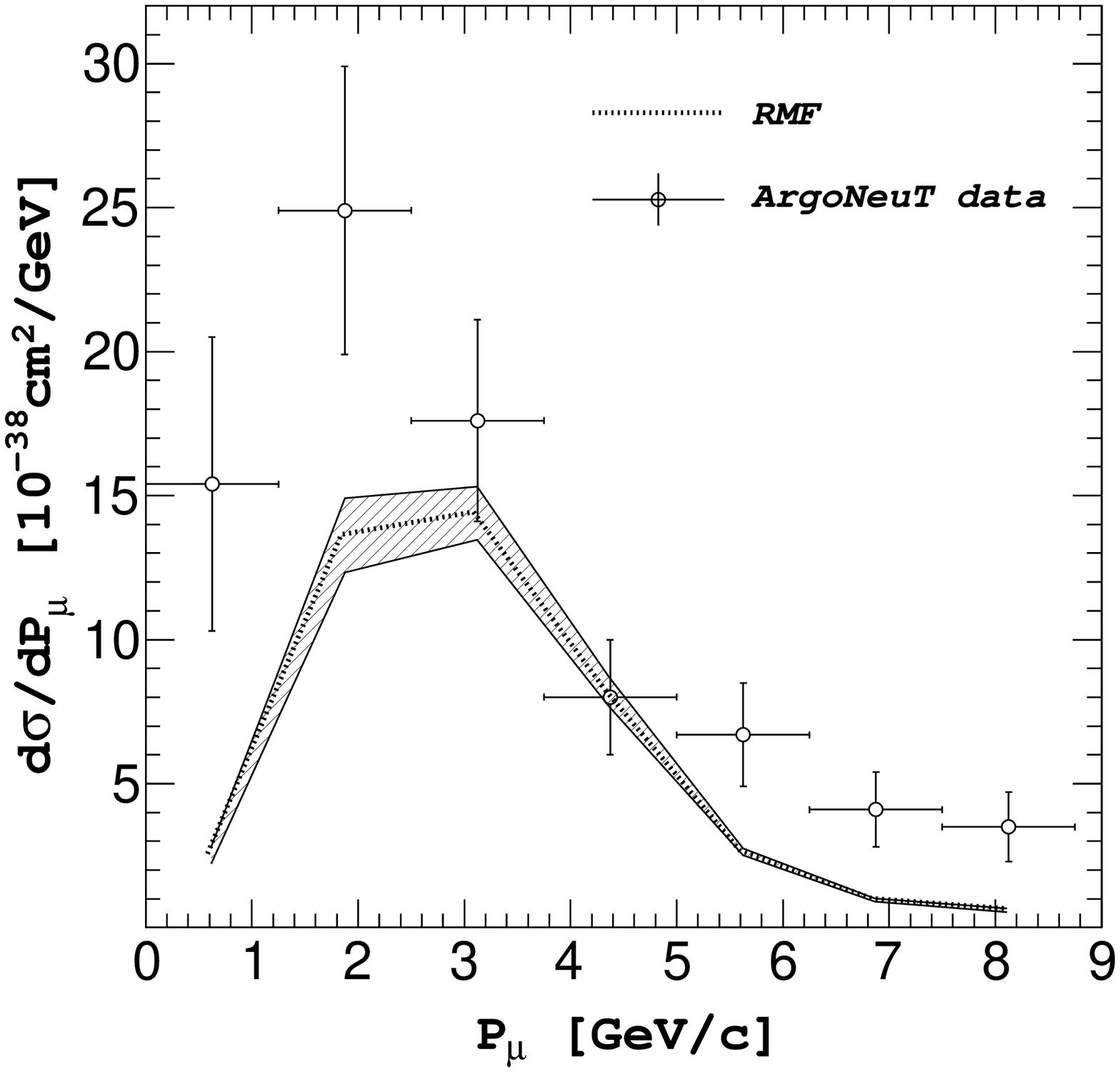} 
\vskip -0.2cm
\caption{(Color online) The same as in Fig. \ref{csrpwia-area} but in RMF.
\label{csrmf-area} }
\end{figure}

In all the calculations presented in this work the bound nucleon states 
are taken as self-consistent Dirac-Hartree solutions derived 
within a relativistic mean field approach using a Lagrangian containing 
$\sigma$, $\omega$, and $\rho$ mesons~\cite{Serot:1984ey}, with medium dependent 
parameterizations of the meson-nucleon vertices that can be more directly related to the
underlying microscopic description of nuclear 
interactions \cite{Finelli:2003fk, Finelli:2004kd, Finelli:2005ni}. 
The same relativistic mean field approach has been used to calculate the 
bound state wave functions in 
\cite{esotici1,Giusti:jp,esotici2}, where the cross sections of the exclusive 
quasi-free $\eep$ and inclusive QE $\ee$ reactions have been 
presented and discussed for oxygen and calcium isotopic chains. 

In the RGF calculations we have used three parameterizations for the 
relativistic OP of $^{40}$Ar: the Energy-Dependent and A-Dependent EDAD1
(where the $E$ represents the energy and the $A$ the atomic number) OP of \cite{Cooper:1993nx}
and  the more recent Democratic (DEM) and the Undemocratic {(UNDEM)} 
phenomenological OPs of \cite{Cooper:2009}.
We note that all these three parameterizations are global ones, since  
they are obtained through a fit to elastic proton-scattering data on a wide
range of nuclei and, as such, they  depend on the atomic number $A$  and  
are not constructed to reproduce the $^{40}$Ar phenomenology.

In Fig. \ref{csteta} the CC differential 
cross section  $d\sigma/ d\theta$ 
integrated
over the ArgoNeuT flux is shown as a function of the muon scattering angle 
$\theta$.
All the calculations give results in reasonable agreement with the
experimental shape but generally underpredict the magnitude of the experimental  
cross section. We note that in the RPWIA FSI are completely 
neglected, while in the RMF the same strong 
energy-independent real potential is used for bound and scattering states. 
The comparison between the RGF results obtained with the EDAD1, DEM, and UNDEM 
potentials can give an idea of how  the predictions of the model are affected 
by uncertainties in the determination of the phenomenological OP. 
The differences depend on the energy and momentum transfer and are 
essentially due to the different imaginary part of the three 
potentials, which accounts for the overall effects of inelastic channels 
and is not univocally
determined from the elastic phenomenology.
In contrast, the real term is similar for different parameterizations and 
gives similar results. We observe that the DEM and UNDEM potentials give in
general close results; in fact, even if they are obtained using very different
fitting procedures \cite{Cooper:2009}, they are based on the same dataset of 
elastic-scattering data.
In constrast, the EDAD1 potential produces somewhat smaller results and larger
differences in comparison with the results of the DEM and UNDEM  potentials:  
for instance, in the peak region with $\theta \approx 5-7$ deg, 
the RGF-DEM and 
RGF-EDAD1 cross sections differ by about $20\% - 30\%$.
   
The RGF cross sections are generally larger than the RPWIA and the RMF ones, 
but they are in general significantly lower than the data, but for 
$\theta \leq 5^{}$ deg and $\theta \geq 20^{}$ deg.
In the RGF the imaginary part of the optical potential redistributes the 
flux in all the final-state channels and,  in each channel, the flux lost 
towards other channels  is compensated by the flux gained from the other
channels. 
The larger cross sections in the RGF arise from the translation to the inclusive 
strength of the overall effects of inelastic channels which are not included 
in the other models such as, for instance, rescattering processes 
of the nucleon in its way out of the nucleus,  non-nucleonic $\Delta$  
excitations which may arise during nucleon propagation, or also some multinucleon 
processes. These contributions are not included explicitly in the RGF, but they 
can be recovered, to some extent, by the imaginary part of the
phenomenological  OP. 

We note that in all the calculations presented in this work we have used 
the standard value of the axial mass $M_A = 1.03$ GeV. A larger value of $M_A$ would 
increase the
cross section and improve the agreement with data.

In Fig. \ref{cspmu} the CC differential 
cross section  $d\sigma/ dP_\mu$ 
integrated
over the ArgoNeuT flux is shown as a function of the muon momentum $P_\mu$.
Also in this case the RPWIA and the RMF results are lower than the 
experimental data, while the RGF produces
larger cross sections which are in better agreement with the data. Anyway, the 
first two measurements in the low energy bins of $P_\mu$ are underestimated by
all the results by a factor of $2$.

All the models which have been adopted for the present calculations are 
based on the IA, where the cross section is given by an incoherent sum of 
interactions between the incident neutrino and all the single nucleons of 
the nucleus.
Also the RGF, which is a more complex model and, with the use of the complex OP, 
is able to recover contributions of channels which are not included in the 
other models, is essentially based on the IA. Models based on IA have been 
successful in the description of QE exclusive and inclusive electron 
scattering \cite{book}. Moreover, the RGF is able to give a reasonable 
description of CCQE and NCE MiniBooNE cross sections 
\cite{Meucci:2011vd,Meucci:2011nc,Meucci:ant}. In the inclusive CC neutrino-nucleus 
scattering at energies larger than a few GeVs, however, all these models may 
neglect important contributions of reaction processes which can be included 
in the experimental cross sections.
It is therefore not surprising that the calculated cross sections in 
Figs. \ref{csteta} and \ref{cspmu} are generally lower than the experimental 
data.

\begin{figure}
\includegraphics[scale=.37]{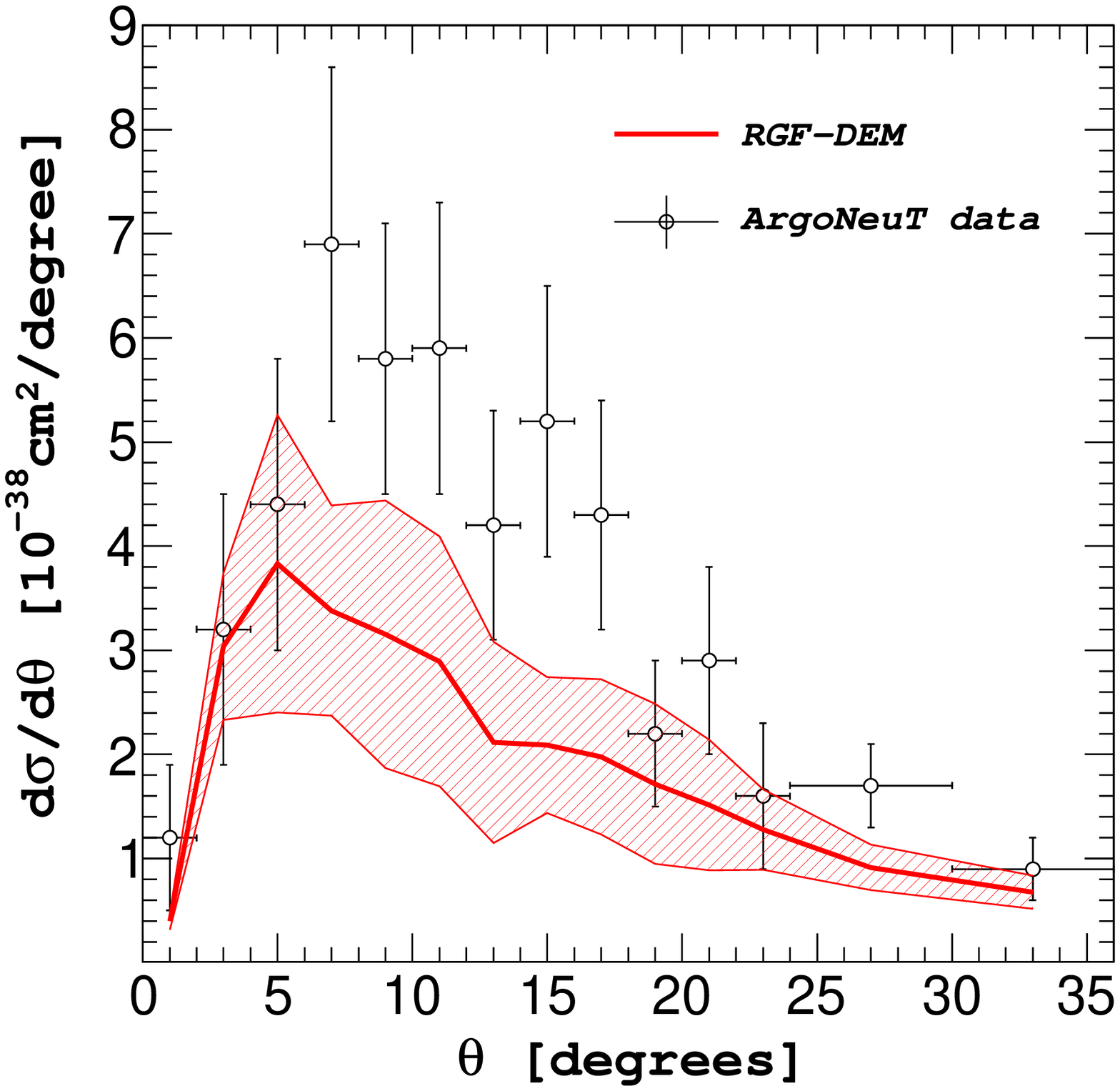} \\
\includegraphics[scale=.37]{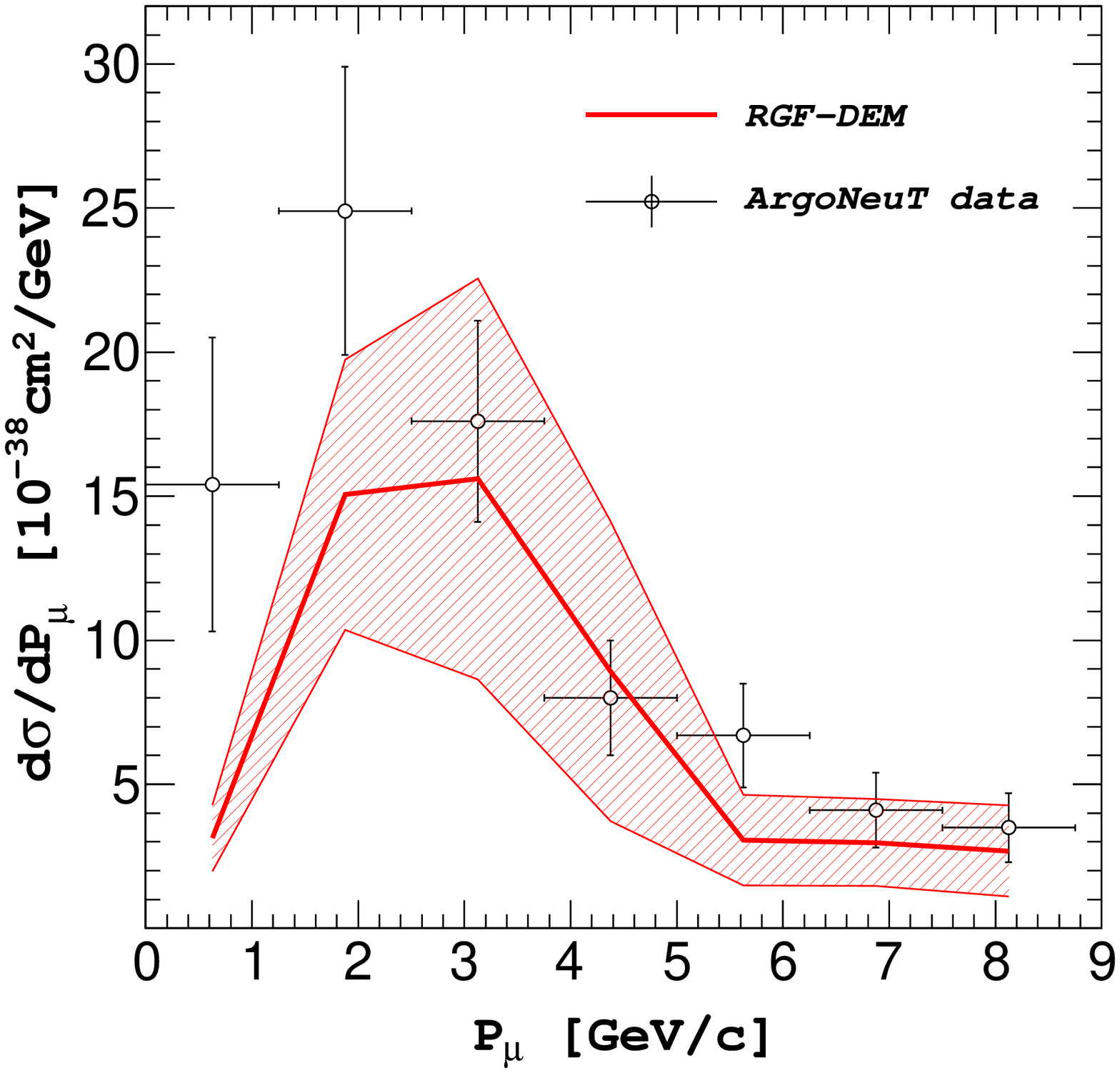} 
\vskip -0.2cm
\caption{(Color online)  The same as in Fig. \ref{csrpwia-area} but in RGF-DEM.
\label{csrgf-area-dem} }
\end{figure}
 
With the aim to give a more quantitative information we have tried to estimate the 
uncertainties of our calculations.
The most obvious source of uncertainty is the neutrino flux: it is usually known with 
sufficient precision but its errors are not negligible \cite{argoneut}. 
For energies up to $10$ GeV the ArgoNeuT flux is 
given in bins of resolution of $1$ or $2$ GeV and, for each bin, we compute 
an \lq\lq average\rq\rq cross section starting from five or more
calculations at different energies. 
It is straightforward to assume that this procedure introduces additional
uncertainties {in} our results. 
In Figs. \ref{csrpwia-area} and \ref{csrmf-area} we show our results for the
$d\sigma/ d\theta$  and $d\sigma/ dP_\mu$ differential 
cross section in RPWIA and in RMF. The bands in the figures represent these two errors added in quadrature. 
In the interval of $\theta$ and $P_\mu$ covered by the ArgoNeuT kinematics the two errors 
are generally small and neither the RPWIA nor the RMF
cross sections can reproduce the experimental data within the error
bands. This is a reasonable result since the RPWIA and RMF cross sections in each bin are stable and the
uncertainties on the neutrino flux are generally small.

In the case of the RGF we consider two additional sources of errors.
The RGF results presented here contain the contribution of both
terms of the hadron tensor in Eq. (25) of \cite{Meucci:2003cv}.
The calculation of the second term, which is entirely due
to the imaginary part of the OP, is a hard and time consuming 
numerical task which requires the integration over all the eigenfunctions of the 
continuum spectrum of the optical potential. Numerical uncertainties on this
term are anyhow under control and, from many calculations in different 
kinematics, have been estimated at most within $10\%$.

The fact that in actual RGF calculations we have to use a phenomenological 
energy dependent OP introduces
additional complications. The OPs in  \cite{Cooper:1993nx,Cooper:2009} have been 
tested for energies up to $\approx 1$
GeV and to provide results up to $\approx 10$ GeV we have to extend the range of validity of
 these parameterizations. This has been done multiplying each term of the 
 OP by a realistic function of the energy that has been chosen to carefully
reproduce the behavior of the OP around $1$ GeV. 
We have checked that our results do not depend significantly on  the multiplying function.
The RGF-DEM 
$d\sigma/ d\theta$  and $d\sigma/ dP_\mu$ differential cross sections are
shown in Fig. \ref{csrgf-area-dem}, where the error bands represent all the  
uncertainties that we have considered added in quadrature. Similar results 
with similar uncertainties are obtained in the case of RGF-UNDEM
and RGF-EDAD1. 
The error bands for the RGF results are larger than for RPWIA and RMF 
and, as a consequence, the upper limits of the RGF cross sections 
are closer to the data. This outcome can be ascribed to the moderately 
large uncertainties on
the cross sections in each experimental bin of neutrino energy, in particular 
for neutrino energies 
of $\approx 2 - 3$ GeV and small energy transferred to the nucleus.

The large error bands in Fig. \ref{csrgf-area-dem} do not allow 
us to draw any definite conclusion. However, the results presented in this work
indicate that the RGF, as well as the other models based on the 
IA, generally underpredict ArgoNeuT data, but they are able to reproduce the shape and 
the correct order of magnitude of the 
experimental cross section. 
 
\section{Summary and conclusions}

In this paper we have compared the predictions of different relativistic descriptions 
of FSI for CC neutrino-nucleus scattering in the ArgoNeuT kinematics.
In the RPWIA  FSI are neglected; in the RMF they are described using the 
same relativistic mean field potential considered in 
describing the initial nucleon state; in the 
RGF the full complex OP, with its real and imaginary parts, is used to
account for FSI. All final-state channels are included in the RGF, the flux lost in each channel is
recovered in the other channels  by the imaginary part
of the OP making use of the dispersion relations
and the total flux is conserved. 
The RGF gives a good description of the $\ee$ data  in the QE region 
and it is also able to describe both CCQE and NCE MiniBooNE 
data. In the RGF cross sections  the contribution of reaction 
channels that are neglected in the other models, e.g., rescattering processes 
of the nucleon in its way out of the nucleus,  non-nucleonic $\Delta$  excitations, which
may arise during nucleon propagation, with or without real
pion production, or also multinucleon processes, is translated, to some extent, 
into the inclusive strength  by the imaginary part of the
phenomenological OP. However, the role of the
various reaction channels included in the phenomenological OP, cannot be 
disentangled and the enhancement of the cross section cannot be attributed 
to a specific process.  
In order to clarify this point, a careful evaluation of all nuclear effects 
and of the relevance of multinucleon emission and of some non-nucleonic
contributions, as well as a  better determination of the 
relativistic OP, which closely fulfills the dispersion relations, would reduce 
the theoretical uncertainties on the RGF. 

Our results give a clear indication that  IA-based models are able to reproduce the
correct order of magnitude and the shape of the ArgoNeuT data but 
they generally underpredict the experimental cross sections, in particular 
for lower values of $P_\mu$ and for values of $\theta$ between $5^{}$ deg and 
$20^{}$ deg. A careful evaluation of all nuclear effects is required to 
recover some important contributions to the CC inclusive strenght. 
In particular, a careful 
study of medium effects in the few-GeV energy region that takes into account 
quasielastic, inelastic, as well as deep-inelastic processes, is highly desirable.

\begin{acknowledgments}

This work was partially supported by the Italian MIUR 
through the PRIN 2009 research project.

\end{acknowledgments}

%

\end{document}